\documentclass[twocolumn]{apex}

\usepackage{amsmath,amsthm,amssymb,mathrsfs}
\usepackage{graphicx} \usepackage{bm} \usepackage{amsmath}
\usepackage{amssymb}

\title{Magnetization switching by microwaves synchronized in the vicinity of precession frequency}

\author{
  Tomohiro Taniguchi} 
\inst{
  National Institute of Advanced Industrial Science and Technology (AIST), Spintronics Research Center, Tsukuba 305-8568, Japan.
}

\abst{
We propose a theoretical framework of a magnetization switching induced solely by a microwave. 
The microwave frequency is always close to but slightly different from 
the oscillation frequency of the magnetization. 
By efficiently absorbing energy from the microwave, 
the magnetization climbs up the energy landscape to 
synchronize the precession with the microwave. 
We introduced a dimensionless parameter $\epsilon$ determining the difference 
between the microwave frequency and the instant oscillation frequency of the magnetization. 
We analytically derived the condition of $\epsilon$ to switch the magnetization, 
and confirmed its validity by the comparison with numerical simulations. 
}
\begin{document}

\maketitle


Microwave-assisted magnetization reversal 
\cite{bertotti01,thirion03,denisov06,sun06,nozaki07,zhu08,bertotti09book,okamoto08,okamoto12,tanaka13} 
is a fascinating subject in magnetism 
for practical applications such as high-density recording. 
An oscillating field generated by microwaves excites a small-amplitude oscillation 
of the magnetization in the ferromagnet 
and significantly reduces the switching field to, typically, 
half of the uniaxial anisotropy field for an optimized microwave frequency. 
However, zero-field switching has not been reported experimentally; 
this is an outstanding problem in this field. 
A proposal of a theoretical possibility for switching 
induced solely by microwaves, 
as well as a deep understanding of its physical picture, 
will be an important guideline 
for further development in this field. 


In previous works on microwave-assisted magnetization reversal 
\cite{bertotti01,thirion03,denisov06,sun06,nozaki07,zhu08,bertotti09book,okamoto08,okamoto12,tanaka13}, 
the microwave source is isolated from the ferromagnet. 
Recently, however, an alternative system has been investigated both experimentally and numerically \cite{suto14,kudo14} 
in which a spin torque oscillator (STO) is used as the microwave source. 
An oscillating dipole field emitted from the STO acts as microwaves on the ferromagnet 
and induces switching. 
Simultaneously, the dipole field from the ferromagnet changes 
the oscillation angle, as well as the oscillation frequency, in the STO. 
Therefore, in this situation, the microwave frequency from the STO depends on 
the magnetization direction in the ferromagnet. 
This motivated us to investigate the possibility of switching the magnetization solely by microwaves, 
the frequency of which depends on the magnetization direction itself. 
The purpose of this letter is to propose a theoretical framework for the switching process. 





\begin{figure}
\centerline{\includegraphics[width=1.0\columnwidth]{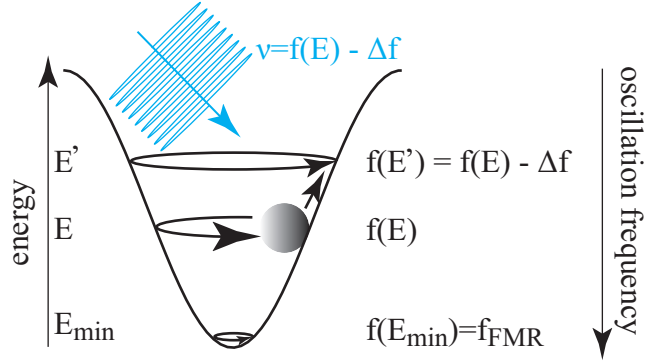}}
\caption{
         Schematic view of the energy landscape of a ferromagnet. 
         When the microwave frequency $\nu$ is close to but slightly different from the oscillation frequency $f(E)$, 
         the magnetization moves on the energy landscape to synchronize these frequencies. 
         \vspace{-3ex}}
\label{fig:fig1}
\end{figure}

Figure \ref{fig:fig1} schematically shows the energy landscape of a ferromagnet. 
The magnetization direction from the stable state is characterized by the energy $E$. 
When the magnetization arrives at the position at which the energy is $E$, 
the magnetic field excites precession of the magnetization on the constant energy curve 
with an oscillation frequency $f(E)$. 
The FMR frequency, $f_{\rm FMR}$, corresponds to $f(E)$ at the minimum energy state. 
Let us assume that the microwave frequency $\nu$ is close to but slightly different from 
the instant precession frequency $f(E)$; 
i.e., $\nu=f(E)-\Delta f$ with $|\Delta f/f| \ll 1$. 
Because the instantaneous oscillation frequency $f(E)$ changes with time during the magnetization dynamics, 
the microwave frequency $\nu$ should also change with time. 
Then, by efficiently absorbing energy from the microwaves, 
the magnetization moves to another constant energy curve $E^{\prime}$ satisfying $f(E^{\prime})=\nu$ 
to synchronize the precession with the microwaves. 
If this shift of the magnetization proceeds toward a higher energy state and occurs continuously, 
the magnetization finally climbs up the energy landscape and switches to the other stable state. 
We introduce a dimensionless parameter $\epsilon$ to characterize the difference 
between the microwave frequency $\nu$ and the instantaneous oscillation frequency $f(E)$. 
An analytical calculation of the energy change with time implies 
a necessary condition of $\epsilon$ for switching. 
The validity of the analytical calculation and the occurence of switching 
induced solely by microwaves are confirmed by a numerical simulation. 



We first describe the system under consideration. 
The magnetization dynamics in a ferromagnet is described by the Landau-Lifshitz-Gilbert (LLG) equation 
\begin{equation}
  \frac{d \mathbf{m}}{dt}
  =
  -\gamma
  \mathbf{m}
  \times
  \mathbf{H}
  -
  \alpha
  \gamma
  \mathbf{m}
  \times
  \left(
    \mathbf{m}
    \times
    \mathbf{H}
  \right),
  \label{eq:LLG}
\end{equation}
where $\mathbf{m}$ is the unit vector pointing in the direction of the magnetization. 
The gyromagnetic ratio is denoted as $\gamma$. 
The second term on the right-hand side of Eq. (\ref{eq:LLG}) is damping with the damping constant $\alpha$. 
The magnetic field $\mathbf{H}$ is related to the magnetic energy density 
$E$ via $\mathbf{H}=-\partial E/\partial (M \mathbf{m})$, 
where $M$ is the saturation magnetization. 
The explicit form of the magnetic field in the present system is given by 
\begin{equation}
  \mathbf{H}
  =
  H_{\rm ac} 
  \cos\psi
  \mathbf{e}_{x}
  +
  H_{\rm ac}
  \sin\psi
  \mathbf{e}_{y}
  +
  H_{\rm K}
  m_{z}
  \mathbf{e}_{z},
  \label{eq:field}
\end{equation}
where we assume that the ferromagnet has uniaxial anisotropy along the $z$-axis 
with the anisotropy field $H_{\rm K}$. 
The microwave amplitude is denoted as $H_{\rm ac}$. 
In the absence of microwaves, the ferromagnet has two stable states given by $\mathbf{m}=\pm \mathbf{e}_{z}$. 
In the following, we assume that the magnetization initially points in the positive $z$-direction, for convention. 
When the microwave frequency $f$ is constant, 
the phase $\psi$ is related to $f$ via 
\begin{equation}
  \psi
  =
  2\pi ft. 
  \label{eq:phase_microwave}
\end{equation}
The torque due to the anisotropy field described by the first term on the right hand side of Eq. (\ref{eq:LLG}), 
$-\gamma \mathbf{m} \times H_{\rm K} m_{z} \mathbf{e}_{z}$, 
describes the precession of the magnetization on a constant energy curve of $E=-MH_{\rm K} m_{z}^{2}/2$. 
The precession frequency on this constant energy curve of $E$ is given by 
\begin{equation}
  f(E)
  = 
  \frac{\gamma}{2\pi} 
  H_{\rm K} 
  m_{z}. 
  \label{eq:resonant_frequency}
\end{equation}
Note that $f(E)$ decreases with increasing energy. 
In other words, the oscillation frequency of the magnetization decreases 
as the magnetization climbs up the energy landscape. 
The FMR frequency is given by $f_{\rm FMR}=\lim_{E \to E_{\rm min}}f(E)=\gamma H_{\rm K}/(2\pi)$. 
Then, in analogy to Eq. (\ref{eq:phase_microwave}), 
we consider the phase of the microwave as given by the relation 
\begin{equation}
  \psi
  =
  \gamma H_{\rm K}
  \left(
    m_{z}
    +
    \epsilon
  \right)
  t.
  \label{eq:phase}
\end{equation}
The microwave frequency will be 
\begin{equation}
  \nu
  \equiv
  \frac{1}{2\pi}
  \frac{d \psi}{dt}
  =
  \frac{\gamma}{2\pi} 
  H_{\rm K}
  \left(
    m_{z}
    +
    \epsilon
    +
    \frac{d m_{z}}{dt}
    t
  \right).
  \label{eq:frequency_model}
\end{equation}
Here, we introduce a parameter $\epsilon$. 
The difference between the microwave frequency 
and the oscillation frequency of the magnetization is 
$\nu - f(E) = \gamma H_{\rm K} [ \epsilon + (dm_{z}/dt)t]/(2\pi)$. 
In particular, when the magnetization precesses on a constant energy curve of $E$, as in the case of resonance, 
$\nu-f(E)=\gamma H_{\rm K} \epsilon/(2 \pi)$. 
Thus, the parameter $\epsilon$ determines the frequency difference. 
We must note that the term $(dm_{z^{\prime}}/dt)t$ is also necessary, 
as mentioned below. 


Before proceeding to further discussion, 
we comment briefly on the above model. 
As mentioned above, the present model is motivated by a ferromagnet coupled to an STO \cite{suto14,kudo14}. 
To strictly investigate the possibility of switching, 
the coupled LLG equations between the ferromagnet and the STO should be solved \cite{comment}. 
It is, however, difficult to solve such LLG equations analytically because of their complexity. 
The present system might be regarded as a simplified model of this system, instead of solving the coupled equations exactly. 
We assume that any complexity of the coupled equations is attributable to the parameter $\epsilon$. 
For example, $\epsilon$ might be related to the delay of the response (frequency change) of the STO 
due to its finite relaxation time 
and can be changed, for example, by using various materials or device geometries. 
Note that the present model is not restricted to a coupled system between a ferromagnet and an STO. 
The use of an arbitrary wave generator as a microwave source is 
another candidate for the present proposal. 
The introduction of the parameter $\epsilon$ provides a wide variety of dynamics in a model 
and will enable us to characterize experimental results by a limited numbers of parameters. 
Although $\epsilon$ is assumed to be constant here, 
it will be interesting to study a model with time-dependent $\epsilon$. 
In fact, as mentioned below, 
the total frequency difference [$\propto \epsilon+(d m_{z}/dt)t$] should be time-dependent 
for switching. 


Next, we consider the analytical condition of $\epsilon$ required to switch the magnetization. 
The energy of the system should increase with time for switching. 
To study the energy change of the ferromagnet, 
it is useful to use a rotating frame $x^{\prime}y^{\prime}z^{\prime}$, 
where the $z^{\prime}$-axis is parallel to the $z$-axis, 
and the $x^{\prime}$-axis always points in the direction of the microwaves \cite{bertotti09book}. 
The LLG equation in the rotating frame is given by 
\begin{equation}
\begin{split}
  \frac{d \mathbf{m}^{\prime}}{dt}
  =&
  -\gamma
  \mathbf{m}^{\prime}
  \times
  \bm{\mathcal{B}}
  -
  \alpha
  \gamma
  \mathbf{m}^{\prime}
  \times
  \left(
    \mathbf{m}^{\prime}
    \times
    \bm{\mathcal{B}}
  \right)
\\
  &+
  \alpha
  \frac{d \psi}{dt}
  \mathbf{m}^{\prime}
  \times
  \left(
    \mathbf{e}_{z^{\prime}}
    \times
    \mathbf{m}^{\prime}
  \right),
  \label{eq:LLG_rotating_frame}
\end{split}
\end{equation}
where $\mathbf{m}^{\prime}=(m_{x^{\prime}},m_{y^{\prime}},m_{z^{\prime}})$ is the unit vector pointing in the magnetization direction in the rotating frame. 
The magnetic field in the rotating frame is 
\begin{equation}
  \bm{\mathcal{B}}
  =
  H_{\rm ac}
  \mathbf{e}_{x^{\prime}}
  +
  \left(
    -\frac{1}{\gamma}
    \frac{d\psi}{dt}
    +
    H_{\rm K}
    m_{z^{\prime}}
  \right)
  \mathbf{e}_{z^{\prime}}.
  \label{eq:field_rotating_frame}
\end{equation}
The second term on the right-hand side of Eq. (\ref{eq:LLG_rotating_frame}) is the damping in the rotating frame. 
A mathematical analogy between the third term and spin torque was pointed out recently \cite{taniguchi14}. 
We define the energy density in the rotating frame as $\mathscr{E}=-M \int d \mathbf{m}^{\prime}\cdot\bm{\mathcal{B}}$. 
Then, from Eq. (\ref{eq:LLG_rotating_frame}), 
the energy change, 
$d \mathscr{E}/dt=(d \mathbf{m}^{\prime}/dt)\cdot(\partial \mathscr{E}/\partial \mathbf{m}^{\prime}) + (\partial \mathscr{E}/\partial t)$, is described as 
\begin{equation}
\begin{split}
  \frac{1}{\gamma M}
  \frac{d \mathscr{E}}{d t}
  =&
  -\alpha
  \left(
    -\frac{1}{\gamma}
    \frac{d \psi}{d t}
    +
    H_{\rm K}
    m_{z^{\prime}}
  \right)
  H_{\rm K}
  m_{z^{\prime}}
  -
  \alpha 
  H_{\rm ac}^{2}
\\
  &\ \ \ +
  \alpha 
  \left[
    H_{\rm ac}
    m_{x^{\prime}}
    +
    \left(
      -\frac{1}{\gamma}
      \frac{d \psi}{d t}
      +
      H_{\rm K}
      m_{z^{\prime}}
    \right)
    m_{z^{\prime}}
  \right]
\\
  &
  \ \ \ \ \ \ \ \ 
  \times
  \left(
    H_{\rm ac}
    m_{x^{\prime}}
    +
    H_{\rm K}
    m_{z^{\prime}}^{2}
  \right)
\\
  &\ \ \ 
  +
  \frac{1}{\gamma^{2}}
  \left(
    \frac{\partial}{\partial t}
    \frac{d \psi}{dt}
  \right)
  m_{z^{\prime}} .
  \label{eq:dEdt}
\end{split}
\end{equation}
Note here that the microwave amplitude, $H_{\rm ac}$, is usually much smaller than 
the uniaxial anisotropy field $H_{\rm K}$. 
In addition, $\mathbf{m} \simeq \mathbf{e}_{z}$ near the initial state. 
Then, the dominant part of the energy change is given by 
\begin{equation}
\begin{split}
  \frac{1}{\gamma M H_{\rm K}^{2}}
  \frac{d \mathscr{E}}{d t}
  \sim
  &
  \alpha
  \left(
    1
    -
    m_{z^{\prime}}^{2}
  \right)
  m_{z^{\prime}}
  \left(
    \epsilon
    +
    \frac{d m_{z^{\prime}}}{dt}
    t
  \right)
\\
  &
  +
  \frac{1}{\gamma H_{\rm K}}
  \frac{d m_{z^{\prime}}}{dt}
  m_{z^{\prime}},
  \label{eq:dEdt_zeroth}
\end{split}
\end{equation}
where we used Eq. (\ref{eq:frequency_model}) in the derivation. 
We note that $d m_{z^{\prime}}/dt <0$ because we are interested in switching 
from $\mathbf{m}=+\mathbf{e}_{z}$ to $\mathbf{m}=-\mathbf{e}_{z}$. 
Because the energy should increase for switching, 
$\epsilon$ should at least satisfy the following condition near the initial state: 
\begin{equation}
  \epsilon
  >
  0.
  \label{eq:condition_epsilon}
\end{equation}
We note that Eq. (\ref{eq:condition_epsilon}) is roughly derived 
without solving the LLG equation exactly 
and by neglecting the higher-order terms of $H_{\rm ac}/H_{\rm K}$. 
Equation (\ref{eq:condition_epsilon}), nevertheless, implies the possibility of switching the magnetization 
solely by microwaves. 
Regarding the above derivation, 
Eq. (\ref{eq:condition_epsilon}) should be regarded as a necessary 
but not sufficient condition for switching. 
Equation (\ref{eq:condition_epsilon}) also implies that 
the maximum frequency of the STO should be higher than the FMR frequency 
if the coupled system between a ferromagnet and an STO is used to test the present model. 
This is because $\nu$ with a positive $\epsilon$ at $t=0$ is larger than $f_{\rm FMR}$ from Eq. (\ref{eq:frequency_model}). 
The sign of $\epsilon$ should be changed for the switching in the opposite direction 
because $d m_{z^{\prime}}/dt>0$ and $m_{z^{\prime}}<0$ in this case. 
We note that $d m_{z^{\prime}}/dt<0$ also implies another necessary condition 
$H_{\rm ac}>\alpha H_{\rm K}/2$ for switching \cite{sun06a}. 



\begin{figure}
\centerline{\includegraphics[width=1.0\columnwidth]{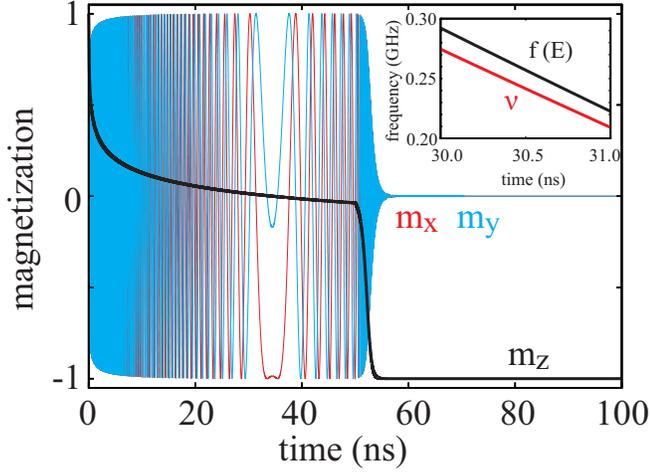}}
\caption{
         Time evolution of the magnetization in the laboratory frame with $\epsilon=0.1$.
         The microwaves are turned off at $t=50$ ns. 
         Inset shows an example of the instantaneous oscillation frequencies of the magnetization $f(E)$ and the microwaves $\nu$. 
         \vspace{-3ex}}
\label{fig:fig2}
\end{figure}



Figures \ref{fig:fig2} shows an example of the magnetization dynamics in the laboratory frame 
obtained by numerically solving Eq. (\ref{eq:LLG}) with $\epsilon=0.1$. 
The values of the parameters are taken from typical experiments and numerical simulations \cite{denisov06,zhu08,okamoto08,okamoto12} as 
$M=1000$ emu/c.c., 
$H_{\rm K}=7.5$ kOe, 
$H_{\rm ac}=450$ Oe, 
$\gamma=1.764\times 10^{7}$ rad/(Oe$\cdot$s), 
and $\alpha=0.01$. 
The initial state is $\mathbf{m}(0)=+\mathbf{e}_{z}$. 
The microwaves are applied starting at $t=0$, and turned off at $t=50$ ns. 
The relaxation dynamics is calculated from $t=50$ ns to $t=100$ ns. 
While the microwaves are applied, 
the magnetization moves from the initial state with precession, 
and finally arrives below the $xy$-plane ($m_{z}<0$). 
After the microwaves are turned off, 
the magnetization relaxes to the switched state, $\mathbf{m}=-\mathbf{e}_{z}$. 
We confirm that $\nu > f(E)$ for $t<0.042$ ns. 
This is consistent with the derivation of Eq. (\ref{eq:condition_epsilon}) that 
$\epsilon + (dm_{z^{\prime}}/dt)t>0$ to increase the energy near the initial state. 
On the other hand, the inset of Fig. \ref{fig:fig2} compares 
the frequencies $f(E)$ and $\nu$, defined by Eqs. (\ref{eq:resonant_frequency}) and (\ref{eq:frequency_model}), respectively, 
after the magnetization moves from the initial state. 
As shown, the instantaneous oscillation frequency of the magnetization, $f(E)$, is always close to but slightly larger than 
the microwave frequency $\nu$, i.e., $f(E) > \nu$. 
Then, by efficiently absorbing energy from the microwaves, 
the magnetization moves to another energy state to synchronize the magnetization precession with the microwaves. 
The point is that this shift of the magnetization corresponds to climbing up the energy landscape. 
Because this shift occurs continuously, 
the magnetization finally arrives at the maximum point of the energy landscape 
and switches its direction. 
The role of the term $(dm_{z^{\prime}}/dt)t$ in Eq. (\ref{eq:frequency_model}) is clarified as follows. 
As mentioned above, $\epsilon$ should be positive to energetically destabilize the initial state. 
However, $\nu-f(E) = \gamma H_{\rm K}[\epsilon + (dm_{z^{\prime}}/dt)t]/(2\pi)$ should be negative 
to climb up the energy landscape by synchronizing $f(E)$ with $\nu$. 
Then, a term such as $(dm_{z^{\prime}}/dt)t$, whose magnitude changes with time, is necessary 
to simultaneously satisfy these two requirements. 



\begin{figure}
\centerline{\includegraphics[width=1.0\columnwidth]{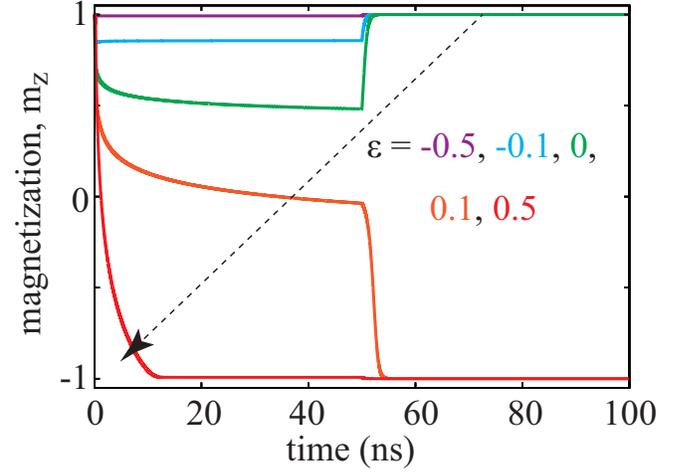}}
\caption{
         Time evolution of $m_{z}$ in the laboratory frame with $\epsilon=-0.5$, $-0.1$, $0$, $0.1$, and $0.5$. 
         The microwaves are turned off at $t=50$ ns. 
         \vspace{-3ex}}
\label{fig:fig3}
\end{figure}



Figure \ref{fig:fig3} summarizes the time evolution of $m_{z}$ for $\epsilon=-0.5$, $-0.1$, $0$, $0.1$, and $0.5$. 
When $\epsilon=-0.5$, $-0.1$, and $0$, 
the magnetization stays close to the initial stable state with a small oscillation amplitude 
because Eq. (\ref{eq:condition_epsilon}) is unsatisfied. 
We note that the minimum of the energy landscape shifts from the $z$-axis due to the microwave field $H_{\rm ac}$ 
pointing in the in-plane direction. 
Therefore, although the magnetization shifts from the $z$-axis in these cases, 
this does not mean that the magnetization is energetically excited. 
On the other hand, when $\epsilon$ becomes $0.1$ and $0.5$, in which Eq. (\ref{eq:condition_epsilon}) is satisfied, 
the ferromagnet absorbs sufficient energy 
and finally arrives at the switched state deterministically. 
These results indicate the possibility of switching solely by microwaves. 
One might consider that there is an upper limit of $\epsilon$ for switching. 
Unfortunately, it is difficult to find such limit from Eq. (\ref{eq:dEdt}) analytically, if it ever exists. 
Instead, we confirmed the switching for $\epsilon \le 100$ numerically.



\begin{figure}
\centerline{\includegraphics[width=0.8\columnwidth]{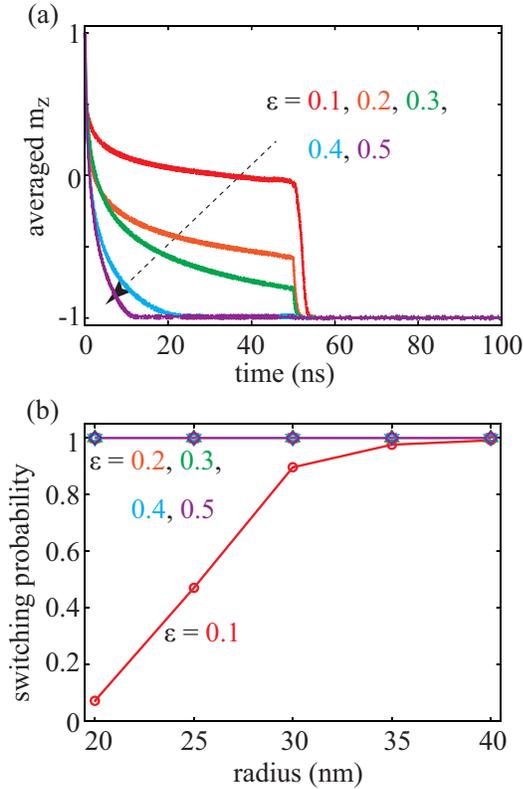}}
\caption{
         (a) Time evolution of $m_{z}$ in the laboratory frame averaged over $N=10^{5}$ samples for $\epsilon=0.1$, $0.2$, $0.3$, $0.4$, and $0.5$. 
             The microwaves are turned off at $t=50$ ns. 
             The radius of the ferromagnet is 35 nm. 
         (b) Dependence of the switching probability on the radius of the ferromagnet. 
             The corresponding thermal stability is $MH_{\rm K}V/(2k_{\rm B}T)=569$ for $r=20$ nm. 
         \vspace{-3ex}}
\label{fig:fig4}
\end{figure}



We also study the effect of thermal fluctuation 
by adding a random torque given by $-\gamma \mathbf{m}\times\mathbf{h}$ to 
the right-hand side of Eq. (\ref{eq:LLG}). 
The component of the random field $\mathbf{h}$ satisfies the fluctuation-dissipation theorem \cite{brown63}, 
$\langle h_{k}(t) h_{\ell}(t^{\prime}) \rangle = [2 \alpha k_{\rm B}T/(\gamma MV)] \delta_{k \ell} \delta(t-t^{\prime})$, 
where the temperature is chosen as the room temperature, $T=300$ K. 
The volume of the ferromagnet is \cite{suto14,kudo14}
$V= \pi r^{2} \times 5$ nm${}^{3}$, 
where $r$ and $d=5$ nm are the radius and thickness, respectively. 
The other material parameters and simulation conditions are identical to those in the above calculations. 
The magnetization dynamics is averaged over $N=10^{5}$ samples \cite{taniguchi12}. 
Figure \ref{fig:fig4} (a) shows the time evolution of the averaged $m_{z}$ 
for $\epsilon=0.1$, $0.2$, $0.3$, $0.4$, and $0.5$. 
The radius is chosen as $r=35$ nm. 
This value makes the cross sectional area almost identical to that in the experiment, 
in which the cross sectional area was an ellipse \cite{suto14}.  
As shown, magnetization switching occurs even in the presence of the thermal fluctuation. 
We also investigate the switching probability at $t=100$ ns, 
which is defined as the number of samples showing $m_{z}(t=100{\rm \ ns})<-0.9$ divided by 
the total number of samples $N=10^{5}$. 
Figure \ref{fig:fig4} (b) shows the relation between 
the switching probability, the parameter $\epsilon$, and the radius $r$. 
When $\epsilon$ is $0.1$, the switching probability is small for small $r$. 
This is because the switching time is relatively long for $\epsilon=0.1$, 
and the thermal fluctuation become large for a small ferromagnet. 
On the other hand, for $\epsilon \ge 0.2$, 
the switching probability at $t=100$ ns is 100 \% 
even in the presence of the thermal fluctuation and for a small volume. 
Therefore, we concluded that switching occurs even in the presence of thermal fluctuation 
when $\epsilon$ is in an appropriate range. 



We emphasize that the present model provides a comprehensive method 
of analytically studying the possibility of switching. 
For example, let us consider autoresonance model \cite{klughertz15} 
in which the microwave frequency of this model is $\nu =f_{0}-at$ 
with constants $f_{0}$ and $a$. 
Because $\nu-f(E)$ should be negative for switching, as mentioned above, 
the constant $a$ should be positive. 
The other switching condition, i.e., 
$d \mathscr{E}/dt$ should be positive near the initial state, 
then requires that $f_{0}>\gamma H_{\rm K}/(2\pi)$. 
These conclusions are consistent with Ref. \cite{klughertz15}. 
The other approach for switching is to restrict $\nu$ to $f(E)$ and neglect the damping \cite{rivkin06}. 
In this case, the magnetization is always in the resonance state. 
Then, from Eq. (\ref{eq:dEdt}), $d \mathscr{E}/dt$ is zero up to the zeroth order of $H_{\rm ac}/H_{\rm K}$. 
This means that the energy change is unnecessary to move from a certain state to the other;
thus, the magnetization can move freely. 
Then, periodic switching between $\mathbf{m}=+\mathbf{e}_{z}$ and $\mathbf{m}=-\mathbf{e}_{z}$ is achieved. 


In conclusion, we proposed a theoretical framework for magnetization switching 
induced solely by microwaves. 
The microwave frequency depends on the magnetization direction 
and is close to but slightly different from the instantaneous oscillation frequency of the magnetization. 
We introduced a dimensionless parameter $\epsilon$ 
that determines the difference between the microwave frequency and the oscillation frequency. 
We analytically derived the necessary condition of $\epsilon$ to switch the magnetization 
from the evolution equation of the energy. 
When $\epsilon$ is in a certain range, 
the magnetization climbs up the energy landscape to 
synchronize the magnetization precession with the microwaves, 
and finally switches its direction. 
We also presented a numerical simulation 
that confirmed the validity of the analytical theory 
and provided evidence of switching. 


The author acknowledges 
Takehiko Yorozu, Shingo Tamaru, Yoichi Shiota, Sumito Tsunegi, Hitoshi Kubota, Koichi Mizushima, Kiwamu Kudo, Osamu Kitakami, and Satoshi Okamoto for valuable discussions. 
In particular, the author is thankful to Kiwamu Kudo for porviding his unpublished data. 
The author also expresses gratitude to Hiroki Maehara, Satoshi Iba, and Ai Emura for their kind encouragement. 
This work was supported by JSPS KAKENHI Grant-in-Aid for Young Scientists (B) 25790044."




\end{document}